# Using Analogical Problem Solving with Different Scaffolding Supports to Learn about Friction

Shih-Yin Lin and Chandralekha Singh

*Department of Physics and Astronomy, University of Pittsburgh, Pittsburgh, PA, 15260 USA*

**Abstract.** Prior research suggests that many students believe that the magnitude of the static frictional force is always equal to its maximum value. Here, we examine introductory students' ability to learn from analogical reasoning (with different scaffolding supports provided) between two problems that are similar in terms of the physics principle involved but one problem involves static friction, which often triggers the misleading notion. To help students process through the analogy deeply and contemplate whether the static frictional force was at its maximum value, students in different recitation classrooms received different scaffolding support. We discuss students' performance in different groups.

## INTRODUCTION

Physics is a subject which contains only a few fundamental principles that are condensed into compact mathematical form. Learning physics requires understanding the applicability of these principles in different contexts which have distinct surface features but involve the same physics. To assist students in recognizing the similarities between different contexts in which the same principle is applicable, various scaffolding mechanisms can be used. For example, students can be taught to perform analogical reasoning between problems that share deep features [1-3].

In this study, we examine introductory physics students' ability to perform analogical problem solving between two problems that are similar in terms of the physics principle involved, but one problem often triggers a misleading notion about static friction which is not applicable in that particular case. Students were asked in a mandatory recitation quiz to first learn from a solved problem provided and take advantage of what they learned from it to solve another problem (called the "quiz problem"). Before solving the quiz problem, students were also explicitly guided to point out the similarities between the two problems. Both problems were about a car in equilibrium on an inclined plane with an inclination of 30 degrees. In the solved problem, the inclined plane was frictionless and the car was held at rest by a rope. The problem asked for the tension in the rope. In the quiz problem, there was no rope present and the car was held at rest by friction. Students were asked to find the frictional force acting on the car. The two problems are similar because the free-body diagrams are analogous and Newton's second law can be applied. The tension in one problem and the friction in the other problem must have the same magnitude. However, prior research [4] suggests that many students believe that the magnitude of the static frictional force ($f_s$) is always equal to its maximum value, the coefficient of static friction ($\mu_s$) times the normal force ($F_N$). This notion is not valid for our quiz problem because the maximum value of static friction exceeds the actual frictional force needed to hold the car at rest.

Prior research [4-5] suggests that students are not necessarily able to exploit the deep analogy between two problems if one of them involves context for which they often have a misleading notion. In order to help students process through the analogy deeply, additional scaffolding supports were provided to students in different interventions as discussed later. Previous research [5] also indicates that students may use Newton's 2nd law to solve for static friction on the car while simultaneously believing that $f_s$ should equal $\mu_s F_N$ (for example, they may first solve for static friction using Newton's 2$^{nd}$ law correctly and then incorrectly calculate $F_N$ by using the equation $f_s = \mu_s F_N$). In order to explore students' thought processes about static friction better and investigate whether the scaffolding support helped, they were asked in the quiz problem to solve for both the static friction and the normal force. Although the solved problem didn't explicitly ask for the normal force, the answer for the normal force can be found in the solution provided.

## METHODOLOGY

313 students from a calculus-based and an algebra-based introductory physics course were recruited in this study (149 and 164 students from each course, respectively). They were divided into three groups - one comparison group and two intervention groups - based on different recitation classes. Students in the comparison group were asked to solve the friction problem in a 15-minute long quiz on their own. No scaffolding support was provided. Students in the two intervention groups received the tension problem as a scaffold to help them solve the friction problem. Additional scaffolding supports were also provided to help them process through the analogy deeply and to carefully contemplate their alternative conceptions about static friction ($f_s = \mu_s F_N$). Depending on the different support provided, different amounts of time were given to students in different groups in order to complete the quiz.

In particular, students in the intervention group 1 were asked to make a qualitative prediction about the magnitude of the static frictional force (whether it's larger or smaller) when the same car is at rest on a steeper inclined plane (with the same coefficient of static friction) based on their daily experience. They were also explicitly instructed to quantitatively calculate the magnitudes of the frictional force acting on the car with two different angles of inclination and compare their quantitative result with their qualitative prediction to check for consistency. We hypothesized that students could reason from their daily experience that it's more difficult to stand still on a steeper inclined plane; therefore, a larger frictional force would be required in order for the same car to stay at rest on a steeper incline. However, if they used $f_s = \mu_s F_N = \mu_s mg \cos\theta$ (where $\theta$ is the angle of inclination) to calculate the magnitude of the frictional force, there would be a conflict because as the angle of inclination increases, the normal force decreases, making the frictional force calculated in this manner smaller. We hypothesized that if students are provided with the solution to the tension problem after noticing this conflict, they could be more likely to notice the deficiency in their original argument. Therefore, students in the intervention group 1 were asked to take the first 10 minutes to do the quiz (which includes extra sub-problems asking for a qualitative prediction and a quantitative calculation of the magnitude of $f_s$ on a steeper incline as well as a consistency check) on their own before the solved tension problem was provided as a scaffolding tool. After they completed the quiz the first time, they turned in their first solution, and then they were given the tension problem with its solution. With the solved tension problem in their possession, they were given 20 minutes to take the quiz a second time.

A different scaffolding support which aimed at guiding students to examine the applicability of the equation $f_s = \mu_s F_N$ was implemented in the intervention group 2. Students in this group were provided with the solved problem and the quiz problem at the same time in a 25-minute quiz. In addition to the instruction which asked them to discuss the similarity between the two problems before solving for the frictional force, they were also asked to explain the meaning of the inequality in $f_s \leq \mu_s F_N$ and discuss whether they can find the frictional force on the car in the quiz problem without knowing $\mu_s$. We intended that this additional questioning provide a direct hint to students to resolve the "conflict" if they are able to recognize the similar roles played by the tension and the friction in the two problems but are concerned about the fact that the equation $f_s = \mu_s F_N$ doesn't yield an answer which has the same magnitude as the tension.

## RESULTS

Before discussing the results we note that students in all groups had adequate time to work on the quiz. In order to examine the effects of different interventions, we investigated how students in different groups approached the friction problem. Tables 1 and 2 list the students' different approaches for finding friction and the corresponding percentage of students in each group in the calculus- and algebra- based course, respectively. As discussed previously, one common mistake students made was to first find the normal force by using Newton's law in the equilibrium situation and then using $f_s = \mu_s F_N$ to solve for friction. We note that if the students' values of the friction force were correct but the overall performance for the whole quiz indicated that they were still connecting the static friction to its maximum value (for example, by using $f_s = \mu_s F_N$ to solve for the normal force in the next sub-problem after finding $f_s = mg \sin\theta$ correctly), they were classified in the $2^{nd}$ category of "$f_s = \mu_s F_N$" in Tables 1 and 2. In addition to this commonly mistaken approach, we found that there were other difficulties students had with the friction problem. For example, some students multiplied $\mu_s$ with a quantity other than the normal force such as the component of the weight parallel to the incline surface. Some students found both the $mg \sin\theta$ and $\mu_s F_N$ terms and set $f_s$ as a combination of them by either adding or subtracting one term to/from the other. There were also students who confused the static friction with the kinetic friction and used $\mu_k$ instead of $\mu_s$ to solve the problem. All these different approaches were placed in the "other" category.

Tables 1 and 2 show that interventions 1 and 2 provided good scaffolding in helping calculus-based students solve the friction problem correctly, while intervention 1 was best in the algebra-based course. The percentage of students who correctly used Newton's 2$^{nd}$ Law in each of these intervention groups was more than two times higher than that for the comparison group in the corresponding course. The Chi-square tests indicate that the differences between these intervention groups and the corresponding comparison groups are significant.

**TABLE 1.** Percentage of students in each group based on their problem solving approaches in the calculus-based course. The boldface font indicates a significant difference (with p-value < 0.05) between the intervention (intv) groups and the comparison group. The *italic font* indicates a marginally significant difference with a p-value between 0.05 to 0.10.

|  | comparison | Intv 1 | Intv 2 |
|---|---|---|---|
| Correct use of Newton's 2nd Law | 21.1 | **56.9** | **56.4** |
| $f_s = \mu_s F_N$ | 42.1 | *25.0* | 30.8 |
| Other | 36.8 | **18.1** | **12.8** |

**TABLE 2.** Percentage of students in each group based on their problem solving approaches in the algebra-based course. The boldface font indicates a significant difference (with p-value < 0.05) between the intervention groups and the comparison group. The *italic font* indicates a marginally significant difference with a p-value between 0.05 to 0.10.

|  | comparison | Intv 1 | Intv 2 |
|---|---|---|---|
| Correct use of Newton's 2nd Law | 14.9 | **60.8** | *28.8* |
| $f_s = \mu_s F_N$ | 34.0 | **15.7** | 37.9 |
| Other | 51.1 | **23.5** | *33.3* |

We note that the increase in the percentage of students who solved the friction problem correctly must be accompanied by a decrease of the number of students who used either $f_s = \mu_s F_N$ or other approaches. Comparing the percentages of students in the "$f_s = \mu_s F_N$" category in particular, however, we found that only intervention group 1 in the algebra-based course showed a significant decrease. Although the percentages in the calculus-based intervention groups 1 and 2 also decreased, the differences from the comparison group (especially that of the intervention group 2) were not large enough to be statistically significant. The finding suggests that although providing students with the solved isomorphic problem gave them more clues about how to construct the problem solution (and therefore the percentages of students in the "other" group and sometimes the "$f_s = \mu_s F_N$" group were reduced), overall, the notion of $f_s = \mu_s F_N$ was still common. A post-activity discussion carried out by the instructor to help more students reconstruct their understanding of static friction may be advantageous.

In general, we found that different interventions had somewhat different effects in helping students adopt a suitable problem solving strategy and avoid common mistakes. Moreover, we observed that calculus- and algebra-based students didn't benefit equally from the same interventions. In the following paragraphs, we discuss students' responses to additional tasks/scaffoldings in different interventions and contemplate the possible reasons why some interventions were more beneficial than others.

As we mentioned earlier, intervention 1 was the best interventions in both the calculus- and algebra-based courses. Students in this intervention group were in particular advised to make a qualitative prediction about the magnitude of the static frictional force on a steeper incline based on their daily experience and compare their prediction with their calculated result. Examining students' answers to these additional questions about the steeper incline when they tried the problem for the first time, we found that most students' reasoning behind their first predictions could be classified into one of three categories: (1) daily experience and correct interpretation/prediction, (2) daily experience and incorrect interpretation/prediction, and (3) answer based on the calculated result. There were students who were able to connect the problem with their daily experience and make a correct prediction. There were also students who knew from their daily experiences that it is less likely for an object to stay at rest on a steeper incline, but the explanations they provided were inconsistent. For example, one student in this category said "Based on my daily experience, the frictional force should be less on a larger incline because it's harder to stay at rest on a steeper incline." We note that the purpose of this prediction question was to help students who originally adopted the $f_s = \mu_s F_N$ approach to discover the conflict between the qualitative trend suggested by the daily experience (static friction should be larger on a steeper incline) and their quantitative answer (showing that the static friction calculated using $f_s = \mu_s F_N$ is smaller). If the students provided alternative explanations about the daily experience as described above, or if they made a prediction not based on their daily experience but based on a quantitative calculation, they were less likely to discover the inconsistency in their responses. In general, we found that these additional questions work in the way we had intended for some students, but not for all of them. Despite this fact, Tables 1 and 2 show that students benefited overall from intervention 1. It is likely that the fact that students in this group had the opportunity to try solving the problem on their own before the solved example was provided is beneficial to them because the clear targeted goal and the thinking process they went through in their first attempt facilitates better transfer to the other problem.

As for intervention 2, which exposed students to the correct inequality $f_s \leq \mu_s F_N$ and asked them to explicitly discuss whether $\mu_s$ is needed to solve the quiz problem by thinking about the meaning of the inequality, the percentage of students who explicitly answered whether $\mu_s$ is needed/not needed is listed in Table 3. By looking at the percentages of students using different problem solving approaches, it appears that this scaffolding support was more beneficial to the calculus-based students than the algebra-based students. Although students were advised to identify the similarity between two problems and were also explicitly asked to think about the fact that the correct expression for the static friction was not $f_s = \mu_s F_N$ but $f_s \leq \mu_s F_N$, more algebra-based students had difficulty in making sense of the inequality and its implication for their static friction problem. As Table 3 shows, fifty percent of the algebra-based students explicitly noted that in order to find the frictional force on the car, $\mu_s$ needs to be given.

**TABLE 3.** Percentage of students in intervention group 2 who answered that $\mu_s$ is needed/not needed in the quiz problem after they explained the meaning of the inequality $f_s \leq \mu_s F_N$.

|  | Calculus | Algebra |
|---|---|---|
| $\mu_s$ not needed | 69.2 % | 45.5 % |
| $\mu_s$ needed | 28.2 % | 50.0 % |
| Irrelevant answer or no answer | 2.6 % | 4.5 % |

Examining students' explanations of the inequality, we found that many algebra-based students weren't able to take advantage of the scaffolding provided because they focused only on one aspect of the inequality and failed to see its full implication. In particular, instead of realizing that "$f_s$ can be any value from zero to the maximum value, which is $\mu_s F_N$, depending on how strong the opposing force is", they only focused on the fact that static friction can't be larger than $\mu_s F_N$. Neither the similarities between the two problems nor explicitly asking them to explain the inequality symbol help them realize that the static friction in the quiz problem is not equal to its maximum value. It is likely that the scaffolding support provided in intervention 2 requires an ability to interpret inequalities at a level which is suitable for calculus-based students but too innovative for many algebra-based students in the framework of preparation for future learning [6]. Therefore, more calculus-based students benefited from the scaffolding provided than the algebra-based students. Future studies involving interviews with students will be conducted to get a more in-depth account of students' thought processes during the task and to exploit the possible strategies to help students.

## SUMMARY AND DISCUSSION

In summary, introductory physics students were able to take advantage of the analogical reasoning activity and transfer their learning from the solved problem provided to solve the quiz problem involving friction if adequate scaffolding support was provided. In particular, intervention 1, in which students had to think before the solved problem was provided was consistently the better intervention in helping students refrain from incorrectly using $f_s = \mu_s F_N$ in both the algebra- and calculus-based courses. This result suggests that providing the solved problem to students only after they have tried to solve the quiz problem on their own was the most beneficial to students in both the calculus- and algebra-based courses.

Moreover, we found that one difficulty students have in learning about the inequality related to friction is that they often focus on static friction not being greater than $\mu_s F_N$, ignoring the fact that $f_s$ can be smaller than this value. In order to help students focus on the inequality better, special effort can be made to address related issues. This analogical reasoning activity, for example, may serve as a good starting point to help students contemplate their understanding of friction if a post-activity discussion is carried out by the instructor. It will be advantageous to discuss with students why the static friction need not always equal its maximum value with the quiz as an example.

In summary, analogical reasoning tasks can provide a good opportunity to help students not only learn about friction, a very challenging topic even at the introductory level, but can also help them build a better knowledge structure. If similar activities and post activity discussions are sustained throughout an introductory physics course, students are likely to develop expertise in physics and become better problem solvers.